\documentclass[aps, prd, 12pt, amsmath, amssymb, superscriptaddress, tightenlines, nofootinbib, letterpaper]{revtex4-1}
\usepackage{amsmath}
\usepackage{slashed}
\usepackage{comment}
\usepackage[pdftex]{graphicx}
\usepackage[pdftex, pdfstartview={FitH}, pdfnewwindow=true, colorlinks=true, citecolor=blue, filecolor=blue, linkcolor=blue, urlcolor=blue, pdfpagemode=UseNone, bookmarks=false]{hyperref}
\usepackage{float}
\usepackage{subcaption}
\usepackage{caption}
\usepackage{xcolor}
\usepackage{soul,ulem}

\usepackage{orcidlink}

\newcommand{\keff}{K_{\mathrm{eff}}}

\DeclareMathOperator{\Pf}{Pf}

\newcommand{\psib}{\overline{\psi}}

\newcommand{\KD}{K\"{a}hler-Dirac }
\newcommand{\Phib}{\overline{\Phi}}

\begin{document}
\title{Anomaly inflow for simplicial \KD fermions}
\author{Simon Catterall}
\affiliation{Department of Physics, Syracuse University, Syracuse, NY 13244, United States}

\begin{abstract}
We study massive \KD fermions propagating on finite triangulations
of various three dimensional manifolds with one or
more boundaries. To probe the anomaly structure of these models
we decompose
the \KD field into two reduced \KD fields and choose boundary conditions that ensure that any edge modes 
are necessarily massless. We then show how to factorize the Pfaffian of this system into a product of bulk and boundary sectors each of which can separately pick up a phase under a global $Z_4$ transformation of the fermions that simultaneously changes the sign of the mass. These phases are topological, depending only on the index of the lattice \KD operator in that sector. For the boundary sector a
non-trivial phase signals the presence of a 't Hooft anomaly. We study cases where the boundary is anomalous but an inflow from the bulk
cancels its phase. Canceling this anomaly for a non-orientable boundary in the absence of the bulk 
recovers the known classification of topological superconductors.
We also give a continuum interpretation of these results in terms of inflow from a gravitational Chern Simons theory
which is induced by integrating out massive bulk \KD fermions. In the presence of a boundary the latter gives rise to Wen-Zee terms that mediate the anomaly inflow mechanism.
\end{abstract}
\maketitle

\newpage

\section{Introduction}
The \KD equation furnishes an alternative to the Dirac equation for describing fermions~\cite{Kahler1962}. The \KD field is composed of a set of antisymmetric tensors or p-forms that are coupled via an action
that involves the exterior derivative and its adjoint. In four flat Euclidean dimensions it is nevertheless equivalent to four degenerate
copies of the Dirac equation with the spinors constructed from the p-forms~\cite{BecherJoos1982,BanksDothanHorn1982}. However, the \KD equation, since it involves tensors and not spinors, differs from the Dirac equation in curved space.  It can be formulated on any manifold without requiring a spin structure~\cite{BennTucker1983} and its global properties are also different
with the zero modes of the massless theory corresponding to harmonic forms~\cite{Rabin1982}. Thus the \KD operator possesses zero modes on compact positive curvature manifolds - something that is impossible
for Dirac fermions~\cite{CatterallLaihoUnmuthYockey2018JHEP,ButtCatterallPradhanToga2021}.

From a lattice perspective it is particularly interesting as it can be discretized without creating fermion doubles~\cite{Rabin1982,BecherJoos1982,BanksDothanHorn1982} -  lattice modes which have no place in the continuum theory and that do not decouple in the continuum limit~\cite{NielsenNinomiya1981}. This renders it attractive in comparison to the Dirac equation~\cite{Catterall2021,Catterall2024}. It also possesses an exact onsite $U(1)$ symmetry which, as we will see, suffers from a mixed anomaly
with gravity even in the discrete theory~\cite{CatterallLaihoUnmuthYockey2018JHEP,ButtCatterallPradhanToga2021,Catterall2023}. Finally, in flat space the theory is equivalent to the well known staggered fermion construction~\cite{KogutSusskind1975,Susskind1977,SharatchandraThunWeisz1981,BanksDothanHorn1982,BecherJoos1982} which has been used very successfully in studies of lattice QCD~\cite{Bazavov2010}.

In this paper we will show that the bulk-boundary correspondence or anomaly inflow
mechanism that is familiar
from the continuum~\cite{CallanHarvey1985,Witten2016RMP,WittenYonekura2021,NguyenSingh2024} also survives in a discrete setting for simplicial \KD fermions
and allows us to re-derive a known mod 2 anomaly~\cite{Catterall2023,CatterallPradhan2024,GuoYou2023} of \KD fermions. Furthermore,
in the continuum we can interpret the inflow as arising from a gravitational Chern Simons theory induced by integrating out massive \KD fermions~\cite{CatterallPradhan2022}.  

Our focus will be on anomaly inflow arising from \KD fermions propagating on a triangulation of a three dimensional bulk geometry
that contains one or more boundary components but in the end it will be easy to generalize the construction to other geometries
and higher dimensions.

\section{\KD fermions on the 2-sphere}\label{sec2d}
We start by reviewing the anomaly for massless \KD fermions propagating on a triangulation of the 2-sphere $S^2$~\cite{CatterallLaihoUnmuthYockey2018JHEP,ButtCatterallPradhanToga2021,Catterall2023}.
The Euclidean massless lattice action is given by
\begin{equation}
    S= \Phib K\Phi=\sum_{p=0}^2\sum_{a=1}^{N_p} \Phib^p_a (K\Phi)^p_a
    \label{action}
\end{equation}
where the lattice or simplicial \KD field $\Phi=(\Phi^0,\Phi^1,\ldots \Phi^D)$ consists of a set of fields $\Phi^p$ living on p-simplices in the triangulation with $\Phib$ independent of $\Phi$ in
Euclidean space. The lattice \KD operator is given by
\begin{equation}
    (K\Phi)^p_a= L_{p,p-1}(a,b)\Phi^{p-1}_b-L_{p,p+1}^\dagger(a,b)\Phi_b^{p+1}
\end{equation}
where $L_{p,p-1}(a,b)$ is an incidence matrix which is only non-zero
when a given $p-1$ simplex $b$ lies in the boundary of the p-simplex $a$ in which case it takes the value $\pm 1$ according to the orientation of the boundary simplex~\cite{Rabin1982,CatterallLaihoUnmuthYockey2018PRD}. 
This action is invariant under the $U(1)_\Gamma$ symmetry:
\begin{equation}
    \Phi\to e^{i\alpha\Gamma}\Phi\quad{\rm and}\quad \Phib\to e^{i\alpha\Gamma}\Phib
\end{equation}
The operator $\Gamma$ distinguishes even simplex fields from
odd simplex fields $\Gamma\Phi^p=\left(-1\right)^p\Phi^p$ and anticommutes with $K$.
If we integrate over the fields we obtain a determinant. Under a $U(1)_\Gamma$ transformation the measure is not invariant
but transforms according to a Jacobian of the form ${\rm det}(J)=e^{2i\alpha{\rm Tr}(\Gamma)}$. Equivalently we retain the original measure and imagine the fermion operator varying as
$K\to e^{i\alpha\Gamma}Ke^{i\alpha\Gamma}$. The partition function
hence transforms
\begin{equation}
    {\rm det}(K)\to e^{2i\alpha\chi}{\rm det}(K)=e^{4i\alpha}{\rm det}(K)
\end{equation}
where we have used the result ${\rm Tr}(\Gamma)=\chi$ the Euler characteristic with $\chi=2$ for the sphere~\cite{Rabin1982,CatterallLaihoUnmuthYockey2018JHEP}. 
This anomaly is linked to the existence of exact zero modes of $K$ which are dictated by the index theorem~\cite{NguyenSingh2024}
\begin{equation}
    {\rm Tr}(\Gamma)=n_+-n_-=\chi
\end{equation}
with $n_\pm$ the number of zero modes with $\Gamma=\pm 1$. The $U(1)_\Gamma$ symmetry is hence broken to $Z_4$ -  the partition function is not invariant under the classical symmetry - there is a mixed $U(1)_\Gamma$-gravitational anomaly - here computed exactly in the lattice
theory~\cite{ButtCatterallPradhanToga2021,Catterall2023}. 

However, the presence of these zero
modes makes the previous discussion rather formal since they force ${\rm det}(K)=0$. One 
can add a (real) mass to lift these zeros leading to
\begin{equation}
    {\rm det}(K+m)=m^\chi\prod^\prime_i (|\lambda_i|^2+m^2)
\end{equation}
This representation shows that the non-zero modes give a real, positive definite contribution to the determinant and the phase is hence carried entirely by the zero modes. 
It is thus topological depending only on the Euler characteristic.
$\chi$. 

Rather than adding a mass
one can absorb these zero modes by inserting an operator with $U(1)_\Gamma$ charge four - a four fermion operator - into the definition of the partition
function so that the result of the fermion integration is ${\rm det}^\prime(K)$ the product
of the non-zero eigenvalues. This operator insertion explicitly breaks the symmetry to $Z_4$.

Since $\{\Gamma,K\}=0$ the massless \KD action can be decomposed into
two independent pieces using the projectors $P_\pm$ where
\begin{equation}
    \Phi_\pm=P_\pm\Phi=\frac{1}{2}(1\pm\Gamma)\Phi\quad{\rm and}\quad \Phib_\pm=P_\pm\Phib
\end{equation}
and the action is
\begin{equation}
    S=\Phib_+ K\Phi_-+\Phib_- K\Phi_+
\end{equation}
Setting say $\Phib_-=\Phi_+=0$ one arrives at a {\it reduced} \KD action
which describes a system with half as many degrees of freedom. This decomposition of a \KD field into reduced fields
is analogous to the decomposition of a Dirac spinor into
Weyl spinors.
If we define
\begin{align}
    a=P_+\Phib^T+P_-\Phi\nonumber\\
    b=P_+\Phi+P_-\Phib^T
\end{align} then the full action can also be written
\begin{equation}
    S=\frac{1}{2}a^TK a+\frac{1}{2}b^T Kb+ma^T\Gamma b
\end{equation} 
In this representation the projection to a single reduced \KD field is equivalent to setting eg. $b=0$. 

To construct a non-zero partition
function for a reduced \KD field requires the insertion of a fermion bilinear with charge two. However, the resultant partition
function will then change sign under a $Z_4$ transformation -- the reduced \KD field propagating on $S^2$ possesses a 't Hooft anomaly. To try to cancel this anomaly we can
try to realize the two dimensional surface as the
boundary of a bulk three dimensional triangulation carrying a massive \KD field. By careful choice
of boundary condition we can then build an anomaly free
model with massless edge modes
and gapped bulk modes in which the
phase arising from a $Z_4$ transformation of the massless boundary modes cancels against a corresponding phase from the bulk partition function. The main thrust of
this paper will be to demonstrate examples of this anomaly inflow mechanism for simplicial \KD fermions.

\section{\KD fermions on the 3-ball $B^3$}
\label{sec3d}
We will consider \KD fermions on a triangulation of the 3-ball $B^3$.
Let us divide the simplices defining this geometry into a set that reside
exclusively on the $S^2$ boundary of the 3-ball which we
label $S$ and the remaining simplices which form
a complementary set in the bulk labeled $B$. To facilitate
analysis we will decompose the field into the
two reduced fields $a$ and $b$ discussed in section~\ref{sec2d}.
Let us further impose the boundary condition that the field $b$ vanishes on $S$:
\begin{equation}
     b_S=0
\end{equation}
In this situation the bulk 
contains one full massive \KD field while the boundary field is a massless reduced \KD fermion.
With the \KD field split into bulk and boundary components $\Phi=(a_B,b_B,a_S)$ the action can be written in block form as
\begin{equation}
 S=\frac12 \Phi^T\mathcal A \Phi,\qquad
 \mathcal A=
 \begin{pmatrix}
 K_B&m\Gamma_B&K_{BS}\\
 - m\Gamma_B&K_B&0\\
 K_{SB}&0&K_S
 \end{pmatrix}.
 \label{eq:Amatrix}\quad{\rm with}\quad K_{SB}=-K_{BS}^T
\end{equation}
The matrix $\cal A$ can then be factorized as ${\cal A}=V^T A V$
\begin{equation}
 \mathcal A=
 \underbrace{\begin{pmatrix}
 1&0\\-C^T A_B^{-1}&1
 \end{pmatrix}}_{V^T}
 \begin{pmatrix}A_B&0\\0&\keff\end{pmatrix}
 \underbrace{\begin{pmatrix}
 1&A_B^{-1}C\\0&1
 \end{pmatrix}}_{V}.
 \label{eq:pfthreefactors}
\end{equation}
where 
\begin{equation}
 A_B=
 \begin{pmatrix}
 K_B& m\Gamma_B\\
 - m\Gamma_B&K_B
 \end{pmatrix}.
 \label{eq:QB}\qquad C=\left(\begin{smallmatrix}K_{BS}\\0\end{smallmatrix}\right)
\end{equation}
and
\begin{equation}
     \keff=K_S+C^T A_B^{-1}C 
\end{equation}
After integration over the fermions one obtains
the Pfaffian of $\cal A$ which takes the factorized form (${\rm det}(V)=1$)
\begin{equation}
 \Pf({\cal A})=\Pf(A_B)\Pf(\keff)
 \label{eq:pfcongruence}
\end{equation}
Substituting for $C$ and $A_B^{-1}=A_B(K_B^2-M^2)^{-1}$ one finds
\begin{equation}
\keff
 =K_S+K_{SB}K_B(m^2-K_B^2)^{-1}K_{BS}.
 \label{eq:Schur}
\end{equation}
At large mass this operator has the expansion
\begin{equation}
 \keff
 =K_S+\frac{1}{m^2}K_{SB}K_BK_{BS}+O(m^{-4})
 \label{eq:largem}
\end{equation}
The presence of
the massive bulk propagator ensures that the lightest modes of the
system are exponentially
localized to the surface in a manner similar to domain wall fermions
\cite{Kaplan:1992bt,Shamir:1993zy}.
Indeed, in the limit of infinite bulk mass the boundary action becomes
that of a reduced
\KD field with the properties
\begin{equation}
 \keff^T=-\keff,\qquad
 \{\keff,\Gamma_S\}=0,\qquad
 \label{eq:properties}
\end{equation}
These results show that no boundary mass is induced via coupling
to the bulk 
for the single boundary field $a_S$. Indeed,
it is not hard to see that
$\keff$, like $K_{S}$, possesses two zero modes. It follows that the boundary Pfaffian, and hence the
free vacuum partition function of the complete 
system vanishes
\begin{equation}
 \Pf\keff=0 \to \Pf({\cal A})=0
 \label{eq:zeroPf}
\end{equation}
As in the discussion in section~\ref{sec2d} this simply instructs us to insert a bilinear
fermion operator to absorb these zero modes in the Grassmann integral and
thereby obtain
a modified partition function which is non-zero. 

However, for the moment
let us ignore these zero modes and examine the result of a $U(1)_\Gamma$ transformation on the fields. This will be useful later when we compare to a continuum perturbative calculation. The reduced components transform as
\begin{equation}a\to e^{i\alpha\Gamma}a\quad{\rm and}\quad b\to e^{i\alpha\Gamma}b\end{equation}
To achieve an invariance of the action we will also assume that the mass can be treated as a spurion field
that also rotates:
\begin{equation}
    m\to m^\prime=e^{-2i\alpha\Gamma}
\end{equation}
One can then think of the fermion operator as transforming ${\cal A}(m)\to e^{i\alpha\Gamma}{\cal A}(m^\prime)e^{i\alpha\Gamma}$ and the corresponding boundary and bulk Pfaffians behave as
\begin{align}
    \Pf(A_B(m))\to e^{2i\alpha\chi_B}\Pf(A_B(m^\prime))\nonumber\\
    \Pf(\keff(m))\to e^{i\alpha\chi_S}\Pf(\keff(m^\prime))
\end{align}
Notice that 
only a single $a$ field contributes to the boundary Pfaffian. The bulk Euler characteristic $\chi_B$ can be found from the relation 
\begin{equation}
    \chi(B^3)=\chi_S+\chi_B=1\to \chi_B=-1\quad {\rm since}\;\chi_S=2
\end{equation}
Thus the phases of both bulk and boundary cancel under a general
$U_\Gamma(1)$ rotation. This is the anomaly inflow mechanism.

Once zero modes are included we know we should consider only
operators that absorb those modes and restrict
ourselves to the $Z_4$ subgroup of $U(1)_\Gamma$. In
this case we see that the modified boundary Pfaffian changes
sign under a $Z_4$ transformation signaling a 't Hooft anomaly.
This is then canceled by a corresponding minus sign arising from the transformation of the bulk Pfaffian.

Notice also that under the $Z_4$ subgroup the mass can be taken real which guarantees the bulk operator can always be inverted. It also means that the spectrum of non-zero modes is paired and, as observed earlier, the anomalous phase arises 
only from the topologically protected zero modes.

\section{\KD fermions on $RP^2\times I$}
\label{secrp2}
It is interesting to ask whether one can construct
similar models where the boundary $S^2$ is replaced by its non-orientable cousin $RP^2$. Since there is no closed 3-manifold whose boundary is $RP^2$ we instead consider the geometry $RP^2\times [0,L]$
which comprises a finite interval terminated
by two $RP^2$ boundaries. Following the logic of section~\ref{sec3d} we choose
boundary conditions that leave one reduced \KD fermion on each boundary. Since the Euler character of the bulk is one it generates
a bulk contribution to the phase which cancels the factor of $-1=i\times i$ coming from the boundaries. We will decompose the field into four
reduced components $F=(a_B,b_B,a_0,a_L)$ where we have set $b_0=b_L=0$ to achieve reduced boundary fermions.
The action can be written 
\begin{equation}
 \bar\Phi(K+m)\Phi=\frac{1}{2}F^T {\cal A} F
\end{equation}
where ${\cal A}$ is given by
\begin{equation}
    {\cal A}=\left(\begin{array}{cccc}K_B&m\Gamma_B&K_{B0}&K_{BL}\\
    -m\Gamma_B&K_B&0&0\\
    -K_{B0}^T&0&K_0&0\\
    -K_{BL}^T&0&0&K_L\end{array}\right)
\end{equation}
Setting
\begin{equation}
    A_B=\left(\begin{array}{cc}K_B&m\Gamma_B\\-m\Gamma_B&K_B\end{array}\right)\quad C=\left(\begin{array}{cc}K_{B0}&K_{BL}\\0&0\end{array}\right)\quad E=\left(\begin{array}{cc}K_0&0\\0&K_L\end{array}\right)\label{CEdefs}
\end{equation}
this can be written
\begin{equation}
    {\cal A}=\left(\begin{array}{cc}A_B&C\\-C^T&E\end{array}\right)
\end{equation}
As before, this admits the Schur factorization ${\cal A}=V^TAV$ where
\begin{equation}
    V=\left(\begin{array}{cc}I&A_B^{-1}C\\0&I\end{array}\right)\quad{\rm and}\quad A=\left(\begin{array}{cc}A_B&0\\0&\keff\end{array}\right)
\end{equation}
with
\begin{equation}
    \keff=E+C^T A_B^{-1} C
    \label{rp2keff}
\end{equation}
Integrating out the fermions once again shows that the Pfaffian of
the full system is simply
\begin{equation}
 \Pf(A_B)\Pf(\keff)
\end{equation}
Plugging eqn.~\ref{CEdefs} into eqn.~\ref{rp2keff} one finds
\begin{equation}
\keff=\left(\begin{array}{cc}{\cal K}_0&D\\
-D^T&{\cal K}_L\end{array}\right)
\end{equation}
where 
\begin{align}
    {\cal K}_0&=K_0+K_{0B}K_B(m^2-K_B^2)^{-1}K_{B0}\nonumber\\
    {\cal K}_L&=K_L+K_{LB}K_B(m^2-K_B^2)^{-1}K_{BL}\nonumber\\
    D&=K_{0B}K_B(m^2-K_B^2)^{-1}K_{BL}\\
\end{align}
In a manner similar to the $B^3$ case the effective boundary operator
is corrected as a result of integrating out the bulk. For large mass, the presence of the bulk propagator ensures that the
light modes are exponentially localized to one of
the two boundaries. These boundaries 
are now coupled via the bulk but the magnitude of
this coupling decreases exponentially with the length $L$.

As in our earlier construction the effective boundary
operator  $\keff$ is antisymmetric, an even function of $m$, and anticommutes with $\Gamma$. It does not behave like a mass term
and does not lift the boundary zero modes. To render the partition function
non-zero one must again insert a single fermion mode on each
boundary. 
Under a $Z_4$ transformation the boundary Pfaffian transforms by a factor $e^{i\pi}=-1$ as one would expect for
two reduced fields living on $RP^2$. Conversely the Pfaffian for the single massive bulk \KD field also transforms by a factor of $-1$ since $\chi_B=1$. Thus the anomalous
phase associated to the boundary fermions is canceled by the phase transformation of the bulk under $Z_4$. Anomaly inflow again. To cancel
the anomaly for each boundary independent of the bulk will require
multiples of four copies of the \KD field.

In the flat space continuum limit each reduced \KD field on the boundary
describes 2 Majorana spinors~\cite{ButtCatterallPradhanToga2021} so canceling this anomaly for each boundary separately would
require multiples of eight Majorana spinors. This agrees 
with the classification of topological superconductors~\cite{FidkowskiKitaev2010,FidkowskiKitaev2011,RyuZhang2012,Qi2013,YaoRyu2013,GuLevin2014,KapustinThorngrenTurzilloWang2015,KaidiParraMartinezTachikawa2020} and the cancellation of
a Dai-Freed anomaly~\cite{DaiFreed1994,Yonekura2016,GarciaEtxebarriaMontero2019} for chiral fermion parity~\cite{DelmastroGaiottoGomis2021,TongTurner2019}. 

\section{Effective bulk continuum action}
\subsection{Manifolds with no boundary}
Since the anomaly is topological and independent of the triangulation it should survive into the
continuum. Indeed if we start with a massive \KD fermion in two dimensions we can argue that the phase of the mass can be rotated
away at the expense of adding a gravitational theta term which is
just the Euler density
\begin{equation}
    \Delta S=i\frac{\theta }{4\pi}\int d^2x \epsilon^{\mu\nu}\epsilon_{ab} R^{ab}_{\mu\nu}
\end{equation}
The anomalous phase encountered under a $U(1)_\Gamma$ transformation then
corresponds to the shift $\theta\to\theta+2\alpha$.
It is then interesting to ask whether there is a corresponding continuum theory that
naturally generates the corresponding three dimensional 
bulk phase.

If we integrate out continuum Dirac fermions in three dimensions in the presence of a background gauge field
we obtain a topological Chern-Simons term~\cite{NiemiSemenoff1983,Redlich1984}. Integrating
out \KD fermions leads to a similar result except that the emergent Chern-Simons term is associated
with a topological gravity theory~\cite{CatterallPradhan2022}. 

To understand how this arises let us first recast the continuum two dimensional \KD field $\Phi=(\phi,\phi_i,\phi_{12})$ in terms of a matrix field $\Psi$~\cite{BecherJoos1982,BanksDothanHorn1982} where 
\begin{equation}
    \Psi=\phi I+\phi_i\gamma^i+\phi_{12}\gamma^1\gamma^2
\end{equation}
where $\gamma^i$ are (Euclidean) Dirac gamma matrices. Each column of this matrix can be interpreted as a Dirac spinor and these spinors can be rotated into each other under a flavor symmetry.
It is then a straightforward exercise to show that the \KD action can be written as
\begin{equation}
    S=\int d^2x\,{\rm Tr}\left[\bar{\Psi}\left(\gamma^i\partial_i+m\right)\Psi\right]
\end{equation}
The matrix fermions carry both Lorentz and flavor indices and this action is
naively invariant under both an (left acting) $SO(2)$ (Euclidean) Lorentz rotation and an (right acting) $SU(2)$ flavor
transformation. However, if we are to retain the original \KD interpretation in terms of p-forms
the Lorentz symmetry must be locked to a $SO(2)$ subgroup of the flavor symmetry with the matrix fermion
transforming as $L\Psi L^\dagger$ under Lorentz transformations. As an aside we remark that the operator $\Gamma$ introduced in the form representation can be realized as $\Psi\stackrel{\Gamma}{\to}\gamma^5\Psi\gamma^5$ and so $\Gamma$ is sometimes called a {\it twisted} chiral symmetry.

Let us see how to generalize this construction to
three dimensions. A three dimensional \KD fermion field consists of eight complex degrees of freedom corresponding to
the set of p-forms in three dimensions ($8=1+3+3+1$). A matrix representation of
this field fermion containing these forms requires using four dimensional gamma matrices
together with a projection that reduces the
total number of degrees of freedom from $16\to 8$.
\begin{equation}
    \Psi_+=P_+\Psi=\frac{1}{2}\left(\Psi+\gamma^5\Psi\gamma^5\right)
\end{equation}
For this projector to commute with $\sum_{\mu=1}^3\gamma^\mu\partial_\mu$ one needs to introduce new 3d gamma matrices
\begin{equation}
    \hat{\gamma}^i=i\gamma^i\gamma^4
\end{equation}
and then write
\begin{equation}
    \Psi_+=\phi I+ \phi_i\hat{\gamma}^i+\frac{1}{2}\phi_{ij}\hat{\gamma}^i\hat{\gamma}^j+\frac{1}{3!}\phi_{ijk}\hat{\gamma}^i\hat{\gamma}^j\hat{\gamma}^k
\end{equation}
where $\Phi=(\phi,\phi_i,\phi_{ij},\phi_{ijk})$ is the 3d \KD field.
For simplicity let us work in Euclidean chiral basis. Then
\begin{equation}
    \hat{\gamma}^i=\left(\begin{array}{cc}\sigma^i&0\\0&-\sigma^i\end{array}\right)\quad{\rm and}\quad \Psi_+=\left(\begin{array}{cc}\psi_+&0\\0&\psi_-\end{array}\right)
\end{equation}
And the massive \KD action reduces to two independent Dirac actions each describing a doublet of
fermions with opposite signs of the mass
\begin{equation}
    \int d^3x\, {\rm tr}\left(\bar{\psi}_+(\sigma^i\partial_i+M)\psi_+-
    \bar{\psi}_-(\sigma^i\partial_i-M)\psi_-\right)
\end{equation}
As a set of Dirac fermions
the system has a internal ${\rm Spin}(4)=SU(2)\times SU(2)$ flavor symmetry corresponding to independent rotations of these doublets plus an additional flavor $U(1)$ symmetry corresponding to $\gamma^5$ which is locked to axial symmetry via the projection. Let us switch on
a background gauge field for
the ${\rm Spin}(4)$ symmetry.  
\begin{equation}
    A_\mu=i\Gamma^a\Gamma^b A_\mu^{ab}+i\Gamma^4\Gamma^a A_\mu^{4a}\quad {\rm where}\;a,b=1\ldots 3
\end{equation}
where $\Gamma^a$ are (Euclidean) Dirac matrices in the flavor space.
In a chiral basis this can be written
\begin{equation}
    \left(\begin{array}{cc}B_\mu&0\\0&\bar{B}_\mu\end{array}\right)
\end{equation}
where 
\begin{align}
    B_\mu&=(\hat{\omega}_\mu^a+e_\mu^a)\sigma^a\nonumber\\
    \bar{B}_\mu&=(\hat{\omega}_\mu^a-e_\mu^a)\sigma^a
\end{align}
where $\hat{\omega}^a_\mu=\frac{1}{2}\epsilon_{abc}A^{bc}_\mu$ and $e^a_\mu=A^{4a}_\mu$. The rationale for the
$(\hat{\omega},e)$ notation will become obvious later.
Integrating out the fermions in the limit $M\to\infty$ yields
an effective action which is the difference of two Chern Simons terms (the relative minus sign reflects the opposite signs of the mass in the two Dirac systems)
\begin{equation}
    \Gamma=ik_+S_{\rm CS}(B)-ik_-S_{\rm CS}(\bar{B})
\end{equation}
where
\begin{equation}
    S_{\rm CS}(B)=\frac{1}{4\pi}\int d^3x\,\epsilon^{\mu\nu\lambda}\left(B_\mu^a\partial_\nu B^a_\lambda+\frac{1}{3}\epsilon^{abc}B_\mu^a B_\nu^b B_\lambda^c\right)
    \label{CSaction}
\end{equation}
and $k_\pm=\frac{1}{2}{\rm sgn}(M).$~\footnote{We use the normalization ${\rm tr}(\sigma_a\sigma_b)=\frac{1}{2}\delta_{ab}$.} 
In fact this can be rewritten keeping the ${\rm Spin}(4)$ invariance explicitly as
\begin{align}
    S_{\rm CS}&=\frac{k}{16\pi}\int d^3x\, \epsilon^{\mu\nu\lambda}\epsilon_{ABCD}\left(A_\mu^{AB}\partial_\nu A_\lambda^{CD}+\frac{2}{3}A_\mu^{AE}A_\nu^{EB}A_\lambda^{CD}\right)\nonumber\\
    \label{CSaction}
\end{align}

As described earlier 
the \KD interpretation of the fermion matrix $\Psi_+$
in terms of p-forms requires that the Lorentz group be locked to a (sub)group of the flavor symmetry. Thus the components of the 
gauge field corresponding to
a ${\rm Spin}(3)$ subgroup of the flavor symmetry can be identified with
the background spin connection $\omega_\mu^{ab}$~\footnote{We assumed a flat background initially but since the answer is topological this can
be generalized to any curved background.}.
The remaining generators of ${\rm Spin}(4)$ transform
as a vector under this ${\rm Spin}(3)$ and are then naturally associated with the background frame $e_\mu^a$~\footnote{Another way of stating this result is that 3d \KD fermions when interpreted as Dirac fermions naturally inherit a flavor structure which, if gauged, gives rise to a Chern Simons theory which can be interpreted as topological gravity.}.
Indeed, the CS action 
written in terms of $\omega$ and $e$ (replacing the latter by $\frac{1}{\ell}e$ to keep
$e$ dimensionless) yields the Einstein Hilbert action with positive cosmological constant in Palatini-Cartan form \cite{Palatini1919,MacDowellMansouri1977}
\begin{equation}
    S=\frac{ik}{4\pi\ell}\int d^3x\,\epsilon^{\mu\nu\lambda} \epsilon_{abc}\left(e_\mu^a R_{\nu\lambda}^{bc}-\frac{2}{3\ell^2}e_\mu^ae_\nu^be_\lambda^c\right)
\end{equation}
where $R_{\nu\lambda}^{bc}$ is the curvature of the spin connection. 
Notice that this is not a generic Einstein-Hilbert action
because the relative coefficients of the two terms are fixed 
so that the action has an hidden ${\rm Spin}(4)$ gauge symmetry corresponding to the Euclidean de Sitter group~\cite{AchucarroTownsend1986,Witten1988,Zanelli2005}.

In summary, we find that
the continuum effective action arrived at after integrating
out a massive \KD fermion on a three dimensional fixed background
space takes the form of a gravitational CS term~\cite{CatterallPradhan2022}. 

\subsection{Manifolds with boundary}
The preceding analysis does not take into account boundaries or interfaces. To understand
anomaly inflow effects 
one needs to prescribe a set of boundary conditions for the gauge fields associated with
such surfaces. To figure out what these should be we can go back
to the fermion description. Let's assume for
simplicity that the 3d geometry in question 
corresponds the half space $z\equiv x_3>0$. This means that the appropriate
fermion mass profile is given by
\begin{equation}
    M(z)=M\theta(z)\quad z\ge 0
\end{equation}
In this background the edge mode solutions take the form $\Psi=f(z)\psi(x)$ where $f(z)=e^{-Mz}$ and
\begin{equation}
    \hat{\gamma}^3\psi=\psi\quad{\rm and}\quad \hat{\gamma}^i\partial_i\psi=0
\end{equation}
These are the usual conditions for Dirac fermions. The \KD character of
the field implies an additional constraint that $\psi\hat{\gamma}^3=\psi$.
Putting these two constraints together we find that $P_-\psi=\frac{1}{2}(\psi-\hat{\gamma}^3\psi\hat{\gamma}^3)=0$ which is just the condition that the
boundary fermion be a reduced \KD field where $\psi_-=\psib_+=0$ as
in the lattice constructions.
The 
only surviving symmetries on the boundary must commute with the boundary
projector $\hat{\gamma}_3$. Because of the locking of
Lorentz to flavor for 
\KD fermions this is the same as $i\Gamma^3\Gamma^4$.
Beside this flavor symmetry the only other good boundary
symmetry corresponds to two dimensional rotations with generator
$i\hat{\gamma}^1\hat{\gamma}^2=i\Gamma^1\Gamma^2$. 
Thus the surviving 
boundary symmetry 
is ${\rm Spin}(2)\times {\rm Spin}(2)$. On $S^2$
this residual symmetry
naturally leads to the following boundary conditions on the gauge fields~\footnote{The vanishing of the fluctuations in the frame fields $e^1_\mu$ and $e^2_\mu$ corresponds to the fact that the boundary is non-dynamical.}.
\begin{equation}
    \omega^{13}_\mu\vert_S=\omega^{23}_\mu\vert_S=e^1_\mu\vert_S=e^2_\mu\vert_S=0
\end{equation}
Notice that the generator of the $U(1)_\Gamma$ symmetry restricted to
the boundary is also $\hat{\gamma}^3$. Thus
\begin{equation}
   \psi\to \hat{\gamma^3}\psi\hat{\gamma^3}=\psi\hat{\gamma}^3=\psi i\Gamma^3\Gamma^4
\end{equation}
where we have employed the edge mode condition
$\hat{\gamma}^3\psi=\psi$. Thus bulk flavor rotations with generator $i\Gamma^3\Gamma^4$ generate the $U(1)_\Gamma$ symmetry
for the boundary fermions. The
corresponding field $A^{34}_\mu=e^3_\mu$ is then a gauge field for this
$U(1)_\Gamma$ symmetry that extends the symmetry into the bulk.
With these boundary conditions the CS action eqn.~\ref{CSaction} (including the factor of i) can be rewritten
\begin{equation}
    S_{CS}=\frac{i}{4\pi}{\rm sgn}(M)\int d^2x\,dz\, \theta(z)\,\epsilon_{\mu\nu\lambda} \left[e^3_\mu R_{\nu\lambda}^{12}+{\rm additional\;terms}\right]
    \label{wenzee}
\end{equation}
The additional terms are similar to the leading term in eqn.~\ref{wenzee} but vanish on the boundary.
The leading piece is a Wen-Zee term~\cite{WenZee1992} and consists
of a coupling of the 2d spin connection $\omega^{12}_\mu$ with the internal $U(1)$ gauge field $e^3_\mu$. Under the 
gauge variation $\delta e^3_z=D_z\alpha$ it yields, after an integration by parts,
\begin{align}
    \delta S_{WZ}&=-\frac{i}{4\pi}{\rm sgn}(M)\int d^2xdz\, \partial_z
    (\theta(z))\alpha \epsilon_{z\nu\lambda} R^{12}_{\nu\lambda}\nonumber\\
    &=-\frac{i}{8\pi}{\rm sgn}(M)\int d^2xdz\,\delta(z)\alpha \epsilon^{\mu\nu}\epsilon_{ab}R_{\mu\nu}^{ab}=-i\alpha\,\chi_{S}\quad M>0
\end{align}
Thus the bulk fermions contribute a phase factor $e^{-i\alpha\chi_{S}}$ to the
partition function under a $U(1)_\Gamma$ rotation. This cancels against the phase $e^{i\alpha\chi_S}$ arising directly from the edge modes which,
as we have seen, corresponds to
reduced \KD fermions. This is the continuum perturbative version of the lattice inflow story described
earlier ignoring boundary zero modes.
Thus the 't Hooft anomaly of the 
massless boundary fermions can be canceled by anomaly inflow from a Chern-Simons theory in the form of a Wen-Zee term associated with
the boundary gauge symmetry~\cite{CatterallPradhan2022}.
Clearly, from the earlier lattice discussion, in order to see the non-perturbative anomaly one needs to also consider non-orientable surfaces.

\section{Conclusions}
We have studied anomaly inflow for
massive simplicial \KD fermions propagating 
in a triangulated three dimensional bulk space equipped
with a boundary. We decompose 
the \KD field into two reduced fields and choose boundary
conditions that ensure that while the bulk fermions
are gapped the edge modes are described by a reduced
field and remain massless. 

On a 3-ball geometry
the edge modes living on $S^2$ experience a 't Hooft anomaly which makes
the associated edge partition function change sign under a 
$Z_4$ transformation. This is compensated
by a minus sign coming from
the determinant of the massive bulk modes giving a concrete
lattice realization of anomaly inflow or the bulk-boundary correspondence.

In the continuum
the local boundary condition that projects the \KD to a
reduced field is sufficient to render the bulk fermion operator self-adjoint since the corresponding
boundary term $\int d^2x\, {\rm tr}\,(\psib\hat{\gamma}^3\psi)$ vanishes. This is to be contrasted with Dirac fermions where a similar local
chiral boundary condition must be replaced by a non-local APS condition for
the bulk operator to be self adjoint. It is the appearance of
this non-local boundary condition that lies at the heart
of the measure problem for chiral fermions. The fact
that a local boundary condition can be employed in the case of \KD fermions is the reason that there is no need to construct an overlap operator for \KD fermions. It is also related to the absence of the eta invariant 
in the Gauss Bonnet identity relating the index of the \KD operator to
the integral of the Euler density as discussed in \cite{NguyenSingh2024}.

It is natural to ask whether what happens when the orientable $S^2$ boundary is replaced by its non-orientable cousin $RP^2$. In fact the
simplest 3-geometry that is anomaly free corresponds to $RP^2\times [0,L]$, with two such $RP^2$ boundaries each carrying a single reduced field. In this case the Pfaffian associated with each boundary changes by a factor of $i$ under
a $Z_4$ transformation giving a net boundary contribution of $-1$ that
is canceled by the sign change of the bulk Pfaffian. 
Clearly to cancel the boundary anomaly completely requires multiples of
four reduced \KD fermions. Locally each such reduced field corresponds to
two Majorana fermions so this cancellation condition matches the usual topological classification of SPT phases~\cite{ButtCatterallPradhanToga2021,Catterall2023,GuoYou2023,AyyarChandrasekharan2016,Catterall2016,Tong2022,WangYou2022}. 

These features are independent of the precise triangulation used
and hence topological in nature. Indeed, we have shown that integrating out massive \KD fermions in the continuum produces an action
that corresponds to a gravitational Chern-Simons theory. In the presence of a boundary 
the gauge symmetry of this Chern Simons theory is broken down to a subgroup which contains
a $U(1)$ symmetry that can be identified with the twisted chiral symmetry of the
boundary. Indeed,
the resultant bulk action then contains a Wen-Zee term that couples this $U(1)$ gauge field to the spin
connection at the boundary~\cite{WenZee1992}. The gauge variation of this
term then
cancels the 't Hooft anomaly of the edge fermions. 

In the continuum theory one can also replace the ${\rm Spin}(4)$ flavor symmetry of the bulk \KD fermions with ${\rm Spin}(3,1)$. If one integrates out the \KD fermions in the corresponding background
gauge field one again obtains a gravitational Chern Simons (CS) theory but now
one whose equations of motion yield (Euclidean) anti-de Sitter space~\cite{AchucarroTownsend1986,Witten1988}. The corresponding $U(1)_\Gamma$
symmetry acting on the boundary fields becomes a
${\rm Spin}(1,1)$ transformation which can be interpreted as a twisted Weyl symmetry acting on the \KD fermions. 

The construction can be easily generalized to higher dimensions.
To achieve naturally massless edge modes one needs to employ boundary conditions that select for reduced \KD fermions. To have a well defined
anomaly inflow problem one needs to balance this requirement
with the topology of the bulk space so that the total
anomaly vanishes. For example the arguments of
section~\ref{secrp2} can be lifted to a geometry $RP^4\times [0,1]$ where applying the same arguments leads to the conclusion that $4k$
copies of the system are needed to cancel the boundary 't Hooft anomaly.
Locally this implies that sixteen Majorana fermions are needed which matches the usual spin-$Z_4$ anomaly cancellation condition.

The corresponding continuum story also generalizes -- integrating out massive
\KD fermions in five dimensions generates a 
5d Chern Simons term whose ${\rm Spin}(6)$ gauge field can be decomposed
into a 5d spin connection and frame corresponding to a 
a topological gravity theory whose symmetry
is the Euclidean de Sitter group \cite{Zanelli2005}. In the
presence of a boundary or interface this CS theory contains a generalization of the Wen-Zee term coupling
a $U(1)$ gauge field for a (twisted) chiral symmetry of the interface fermions
to the four dimensional Euler density.

Clearly, \KD fermions are capable of probing interesting aspects
of geometry. In this paper we have examined how the usual bulk boundary
correspondence or anomaly inflow mechanism works for such fermions. One of the advantages of \KD fermions is that this analysis can be performed directly
on a  finite lattice with no
need to take a continuum limit. One might hope that they can therefore serve
as a useful analog toy model for understanding the chiral fermion problem. In fact the connection to chiral fermions is stronger than this -- if one decomposes
the minimal anomaly free theory of reduced \KD fermions into spinors one
discovers the Pati-Salam GUT \cite{Catterall2023}. More concretely, 
one finds in the
continuum that the global symmetries and matter representations of the theory match those of Pati-Salam - a chiral
gauge theory containing the Standard Model. Furthermore, as
we have seen, it has a natural lattice formulation 
that preserves certain key symmetries and their anomalies.  However a successful approach to (a class of) chiral lattice gauge theories along these lines would 
necessitate gauging certain offsite symmetries - the
shift symmetries of staggered fermions. This is a very non-trivial problem and indeed it is not clear
that a solution is even possible. Nevertheless, first steps in this direction can be found in \cite{CatterallPradhan2024}.

\begin{acknowledgments}
SC was supported by DOE grant DE-SC0009998. He acknowledges helpful discussions with 
Scott Collier, Jay Hubisz, Ho Tat Lam and Alex Maloney. Anthropic's Claude AI was used to help generate references and check algebra.
\end{acknowledgments}

\bibliographystyle{apsrev4-1} 
\bibliography{references} 
\vfill\newpage


\end{document}